\documentclass[proceedings]{JHEP3}

\PrHEP{PrHEP unesp2002}                 
\conference{Workshop on Integrable Theories, Solitons and Duality}

\title{On the Octonionic M-superalgebra}

\author{\speaker{Francesco Toppan}    
\thanks{A large part of the results here reported is a fruit of a collaboration
with J. Lukierski.}\\  
        \\
CBPF - CCP, Rua Dr. Xavier Sigaud
150, cep 22290-180 Rio de Janeiro (RJ), Brazil.

        E-mail: \email{toppan@cbpf.br}}    

\abstract{The generalized supersymmetries admitting abelian
bosonic tensorial central charges are classified in accordance
with their division algebra structure (over ${\bf R}$, ${\bf C}$,
${\bf H}$ or ${\bf O}$). It is shown in particular that in $D=11$
dimensions, the $M$-superalgebra admits a consistent octonionic
formulation, involving $52$ real bosonic generators (in place of
the $528$ of the standard $M$-superalgebra). The octonionic $M5$
(super-$5$-brane) sector coincides with the octonionic $M1$ and
$M2$ sectors, while in the standard formulation these sectors are
all independent. The octonionic conformal and superconformal
$M$-algebras are explicitly constructed. They are respectively
given by the $Sp(8|{\bf O})$ ($OSp(1,8|{\bf O})$) (super)algebra
of octonionic-valued (super)matrices, whose bosonic subalgebra
consists of $232$ (and respectively $239$) generators.}

\begin{document}
 \section{Introduction.}
The generalized supersymmetries going beyond the standard H{\L}S
scheme \cite{hls} admit the presence of bosonic abelian tensorial
central charges associated with the dynamics of extended objects
(branes). Classification schemes for generalized supersymmetries
are now available \cite{fer}. It is worth mentioning that they are
based on previous mathematical classifications \cite{{abs},{kt}}
for spinors and Clifford algebras, in terms of the associative
division algebras (${\bf R}$, ${\bf C}$, ${\bf H}$).\par Recently,
we investigated \cite{{lt1},{lt2}} the possibility of realizing
general supersymmetries in terms of the non-associative division
algebra of the octonions. Our work was the first one to suggest a
possible octonionic version of the $M$-theory, with the explicit
construction of its corresponding $M$-superalgebra. In the past,
algebras of (super)-matrices with octonionic-valued entries have
been introduced in the context of the ten-dimensional superstring
theory \cite{{fm},{cs}}. \par Due to the non-associative character
of the octonions, the octonionic-valued generalized
supersymmetries have peculiar and very surprising features which
will be discussed at length in the following. Perhaps the most
remarkable and the most unexpected of such features consists in
the fact that the different bosonic sectors expressed by the
tensorial abelian central charges are no longer independent, as
for the standard generalized supersymmetries admitting associative
realizations, but they are all inter-related. This phenomenon is a
peculiar characteristic of the octonionic construction.\par It is
worth noticing that the Minkowskian $11$-dimensional spacetime
supports two inequivalent structures, the real structure and the
octonionic one. Therefore, besides the standard $M$-algebra
leading to the $OSp(1|32)$ superalgebra \cite{bv} (and its
$OSp(1|64)$ superconformal algebra), a different $M$-algebra can
be introduced in terms of the octonionic structure and
consistently defined as a closed algebra. This is the octonionic
$M$-algebra (it will also sometimes be referred to as
$M$-superalgebra) which will be discussed in this talk. Of course,
it is too early to say whether this octonionic $M$-algebra can be
of any relevance for physics. On the other hand, the mere fact
that it exists, side by side with the standard $M$-algebra (not to
mention its puzzling features) justifies a thorough investigation
of this and its related mathematical structures.\par The plan of
this talk is as follows. In the next section the classification of
Clifford algebras and spinors (i.e. the necessary ingredients to
introduce supersymmetry) is recalled. Later, in section $3$, the
connection of division algebras with the classification of
Clifford algebras will be elucidated. In particular the
octonionic-valued realizations (which are usually disregarded in
the literature) of the Clifford algebras and their corresponding
spinors will be introduced. This paves the way for the
construction, in Section $4$, of the generalized supersymmetries
based on the division algebras and, in Section $5$, of the
octonionic $M$-algebra. A detailed discussion of its properties
will also be given. In particular a table, based on the octonionic
structure constants, expressing the equivalence of the different
brane sectors in the octonionic description, will be furnished. In
Section $6$ the octonionic superconformal $M$-algebra will be
introduced. Finally, in the Conclusions, the relation of the
octonionic $M$-algebra with other algebraic structures such as
Jordan algebras will be elucidated. Some possible geometrical
interpretations underlining the octonionic description will be
pointed out and the outline for further future investigations will
be given.

\section{Classifying Clifford algebras and spinors.}

The generalized space-time supersymmetries are the ones going
beyond the standard H{\L}S scheme \cite{hls}. This implies that
the bosonic sector of the Poincar\'e or conformal superalgebra no
longer can be expressed as the tensor product structure
$B_{geom}\oplus B_{int}$, where $B_{geom}$ describes space-time
Poincar\'e or conformal algebras and the remaining generators
spanning $B_{int}$ are Lorentz-scalars.\par In the particular case
of the $D=11$ dimensions, where the $M$-theory should be found,
the following construction is allowed. In the $D=11$ Minkowskian
spacetime with signature $(10,1)$ the spinors are real and have
$32$ components.\par By taking the anticommutator of two real
spinors the most general result that we can expect consists of a
$32\times 32$ symmetric matrix, which admits
$32+\frac{32\cdot31}{2}=528$ components. On the other hand, the
standard supertranslation algebra underlining the maximal
supergravity contains only the $11$ bosonic Poincar\'e generators
and by no means the r.h.s. saturates the number of $528$. The
extra generators that should be expected in the right hand side
are obtained by taking the totally antisymmetrized product of $k$
Gamma matrices (the total number of such objects is given by the
Newton binomial ${\textstyle{\left(\begin{array}{c}
  D \\
  k
\end{array}\right)}}$).
The most general $32\times 32$ matrix can be constructed. The
further requirement of being symmetric implies that the total
number of $528$ is obtained by summing the $k=1$, $k=2$ and $k=5$
sectors, so that $528=11+55+462$. The most general supersymmetry
algebra in $D=11$ can therefore be presented as
\begin{equation}
\{Q_a,Q_b\}= (C\Gamma_\mu )_{ab}P^\mu +(C\Gamma_{[\mu\nu]})_{ab}
Z^{[\mu\nu]} +
(C\Gamma_{[\mu_1\dots\mu_5]})_{ab}Z^{[\mu_1\dots\mu_5]}
\label{Malg}
\end{equation}
(where $C$ is the charge conjugation matrix).\par $Z^{[\mu\nu]}$
and $Z^{[\mu_1\dots\mu_5]}$ are tensorial central charges, of rank
$2$ and $5$ respectively. These two extra central terms on the
right hand side correspond to extended objects \cite{{bst},{dk}},
the $p$-branes. The algebra (\ref{Malg}) is called the
$M$-algebra. It provides the generalization of the ordinary
supersymmetry algebra recovered by setting $Z^{[\mu\nu]} \equiv
Z^{[\mu_1\dots\mu_5]}\equiv 0$.
\par
For the purpose of the classification of generalized
supersymmmetries in any given signature space-time, we need at
first to have at disposal the mathematical classification of
Clifford algebras and spinors (see \cite{{abs},{kt}}, while a very
convenient reference for connection with physics is \cite{oku}).
In the rest of this section we introduce the fundamental results
which will be used in the following. Such results can be very
conveniently expressed in terms of the recursive algorithm given
below. Two remarks are in order. The first one: despite the fact
that a quantum theory is described by complex numbers, without
loss of generality (complex numbers can be considered as points in
the real plane) it is preferable to work with Clifford algebras
expressed by real-valued matrices. The structure of Clifford
algebras is much clearer in such a framework (e.g. its connection
with division algebras properties). A second comment: the
algorithm furnished below permits in individuating a single
representative for each irreducible class of representations of
Clifford's Gamma matrices.\par The construction is as follows. Let
us prove at first that a recursive construction of $D+2$ spacetime
dimensional Clifford algebras is available, when assumed known a
$D$ dimensional representation. Indeed, it is a simple exercise to
verify that if $\gamma_i$'s denotes the $d$-dimensional Gamma
matrices of a $D=p+q$ spacetime with $(p,q)$ signature (namely,
providing a representation for the $C(p,q)$ Clifford algebra) then
$2d$-dimensional $D+2$ Gamma matrices (denoted as $\Gamma_j$) of a
$D+2$ spacetime are produced according to either
\begin{eqnarray}
 \Gamma_j &\equiv& \left(
\begin{array}{cc}
  0& \gamma_i \\
  \gamma_i & 0
\end{array}\right), \quad \left( \begin{array}{cc}
  0 & {\bf 1}_d \\
  -{\bf 1}_d & 0
\end{array}\right),\quad \left( \begin{array}{cc}
  {\bf 1}_d & 0\\
  0 & -{\bf 1}_d
\end{array}\right)\nonumber\\
&&\nonumber\\ (p,q)&\mapsto&
 (p+1,q+1).\label{one}
\end{eqnarray}
or
\begin{eqnarray}
 \Gamma_j &\equiv& \left(
\begin{array}{cc}
  0& \gamma_i \\
  -\gamma_i & 0
\end{array}\right), \quad \left( \begin{array}{cc}
  0 & {\bf 1}_d \\
  {\bf 1}_d & 0
\end{array}\right),\quad \left( \begin{array}{cc}
  {\bf 1}_d & 0\\
  0 & -{\bf 1}_d
\end{array}\right)\nonumber\\
&&\nonumber\\ (p,q)&\mapsto&
 (q+2,p).\label{two}
\end{eqnarray}
Some remarks are in order. The two-dimensional real-valued Pauli
matrices $\tau_A$, $\tau_1$, $\tau_2$ which realize the Clifford
algebra $C(2,1)$ are obtained by applying either (\ref{one}) or
(\ref{two}) to the number $1$, i.e. the one-dimensional
realization of $C(1,0)$. We have indeed
\begin{eqnarray}
&\tau_A= \left(\begin{array}{cc}0 &1\\ -1&0  \end{array}\right),
\quad \tau_1= \left(\begin{array}{cc}0 &1\\ 1&0
\end{array}\right), \quad
\tau_2= \left(\begin{array}{cc}1 &0\\ 0&-1  \end{array}\right).
\quad &\label{Pauli}
\end{eqnarray}
All Clifford algebras are obtained by recursively applying the
algorithms (\ref{one}) and (\ref{two}) to the Clifford algebra
$C(1,0)$ ($\equiv 1$) and the Clifford algebras of the series
$C(0, 3+4m)$ ($m$ non-negative integer), which must be previously
known. This is in accordance with the scheme illustrated in the
table below.\par {~}\par {\em Table with the maximal Clifford
algebras (up to $d=256$).}
 {\tiny{
{\begin{eqnarray} &
\begin{tabular}{|cccccccccccccccccccc}
\hline
   1  &$\ast$& 2&$\ast$&  4&$\ast$& 8&$\ast$&16&$\ast$&32&$\ast$&64&
   $\ast$&128&$\ast$&256&$\ast$
      \\ \hline
  &&&&&&&&&&&&&\\
 $\underline{(1,0)}$ &$\Rightarrow$& $(2,1)$ &$\Rightarrow$&(3,2)
 &$\Rightarrow$&
  (4,3) &$\Rightarrow$&(5,4)& $\Rightarrow$ &(6,5)
&$\Rightarrow$&
  (7,6) &$\Rightarrow$&(8,7)& $\Rightarrow$ &(9,8)
  &$\Rightarrow$

  \\
  &&&&&&&&&&\\
  &&&&&&&&&&\\
   &&&&&&(1,4)&$\rightarrow $&(2,5)&$\rightarrow$&(3,6)
&$\rightarrow $&(4,7)&$\rightarrow$&(5,8)&$\rightarrow
$&(6,9)&$\rightarrow$
   \\
  &&&&&$\nearrow$&&&&&\\
  &&&&{\underline{(0,3)}}&& &&&&\\
  &&&&&$\searrow$&&&&&\\
  &&&&&&&&&&\\
  &&&&&&(5,0)&$\rightarrow
  $&(6,1)&$\rightarrow$&(7,2)&$\rightarrow$
&(8,3)&$\rightarrow $&(9,4)&$\rightarrow$&(10,5)&$\rightarrow$
  \\
   &&&&&&&&&&\\
  &&&&&&&&&&\\
   &&&&&&&&(1,8)&$\rightarrow $&(2,9)
&$\rightarrow $&(3,10)&$\rightarrow$&(4,11)&$\rightarrow
$&(5,12)&$\rightarrow$
   \\
  &&&&&&&$\nearrow$&&&\\
  &&&&&&{\underline{(0,7)}}&& &&\\
  &&&&&&&$\searrow$&&&\\
  &&&&&&&&&&\\
  &&&&&&&&(9,0)&$\rightarrow $&(10,1)
&$\rightarrow $&(11,2)&$\rightarrow$&(12,3)&$\rightarrow
$&(13,4)&$\rightarrow$
\\
&&&&&\\ &&&\\
  &&&& &&&&&&&&&&(1,12)&$\rightarrow $&(2,13)
&$\rightarrow $
   \\
  &&&&&&&&&&&&&$\nearrow$&&&\\
  &&&&&&&&&&&&{\underline{(0,11)}}&& &&\\
  &&&&&&&&&&&&&$\searrow$&&&\\
  &&&&&&&&&&&&&&\\
  &&&&&&&&&&&&&&(13,0)&$\rightarrow $&(14,1)
&$\rightarrow $\\ &&&&&\\ &&&\\
 &&&& &&&& &&&&&&&&(1,16)&$\rightarrow $
   \\
  &&&&&&&&&&&&&&&$\nearrow$&&&\\
  &&&&&&&&&&&&&&{\underline{(0,15)}}&& &&\\
  &&&&&&&&&&&&&&&$\searrow$&&&\\
  &&&&&&&&&&&&&&\\
  &&&&&&&&&&&&&&&&(17,0)&$\rightarrow $\\

\end{tabular}&\nonumber
\end{eqnarray}}}}
\begin{eqnarray}\label{bigtable}
&& \end{eqnarray} Concerning the previous table, some remarks are
in order. The columns are labeled by the matrix size ${\bf d}$ of
the maximal Clifford algebras. Their signature is denoted by the
$(p,q)$ pairs. Furthermore, the underlined Clifford algebras in
the table are called the ``primitive maximal Clifford algebras".
The remaining maximal Clifford algebras appearing in the table are
the ``maximal descendant Clifford algebras". They are obtained
from the primitive maximal Clifford algebras by iteratively
applying the two recursive algorithms (\ref{one}) and (\ref{two}).
Moreover, any non-maximal Clifford algebra is obtained from a
given maximal Clifford algebra by deleting a certain number of
Gamma matrices. It should be noticed that Clifford algebras in
even-dimensional spacetimes are always non-maximal.

Let us discuss concretely a given example, namely the explicit
construction of the $D=p+q$ spacetime dimensional Clifford
algebras for $D=11$ (the dimensionality of M-theory). We obtain
\begin{eqnarray}&
\begin{tabular}{|cl|c|}
\hline $(p,q)$& {\em type}&$d$\\ \hline
  (11,0) & $\subset$ (11,2)& 64  \\
  (10,1) &{\bf M}& 32  \\
  (9,2) & $\subset$ (11,2) &64  \\
  (8,3) & {\bf M} &64   \\
  (7,4) & $\subset$ (7,6) &64   \\
  (6,5) &  {\bf M}&32   \\
  (5,6) & $\subset$ (7,6) &64   \\
  (4,7) &  {\bf M}&64   \\
  (3,8) &  $\subset$ (3,10)&64   \\
  (2,9) & {\bf M}&32   \\
  (1,10) & $\subset$ (3,10)& 64   \\
  (0,11) & {\bf M}& 32 \\ \hline
\end{tabular}&\nonumber\end{eqnarray}
where the maximal Clifford algebras are labeled by ${\bf M}$ (the
remaining non-maximal algebras are recovered from the maximal ones
given on the second column, after deleting a certain number of
$\Gamma$-matrices). The size of the matrix representation is given
by the number on the right ($d$). \par For what concerns the
construction of the primitive maximal Clifford algebras of the
series $C(0, 3+8n)$ (also known as quaternionic series, due to its
connection with this division algebra, as we will see later), as
well as the octonionic series $C(0,7+8n)$, the answer can be
provided with the help of the three Pauli matrices (\ref{Pauli}).
We construct at first the $4\times 4$ matrices realizing the
Clifford algebra $C(0,3)$ and the $8\times 8$ matrices realizing
the Clifford algebra $C(0,7)$. They are given, respectively, by
\begin{eqnarray}
C(0,3) &\equiv& \begin{array}{c}
  \tau_A\otimes\tau_1, \\
  \tau_A\otimes\tau_2, \\
  {\bf 1}_2\otimes \tau_A.
\end{array}
\end{eqnarray}
and
\begin{eqnarray}
C(0,7) &\equiv& \begin{array}{c}
  \tau_A\otimes\tau_1\otimes{\bf 1}_2, \\
  \tau_A\otimes\tau_2\otimes{\bf 1}_2, \\
  {\bf 1}_2\otimes \tau_A\otimes \tau_1,\\
  {\bf 1}_2\otimes \tau_A\otimes \tau_2,\\
  \tau_1\otimes{\bf 1}_2\otimes\tau_A,\\
  \tau_2\otimes{\bf 1}_2\otimes\tau_A,\\
  \tau_A\otimes\tau_A\otimes\tau_A.
\end{array}\label{c07}
\end{eqnarray}
The three matrices of $C(0,3)$ will be denoted as ${\overline
\tau}_i$, $=1,2,3$. The seven matrices of $C(0,7)$ will be denoted
as ${\tilde \tau}_i$, $i=1,2,\ldots,7$. \par In order to construct
the remaining Clifford algebras of the series we need at first to
 apply the
(\ref{one}) algorithm to $C(0,7)$ and construct the $16\times 16$
matrices realizing $C(1,8)$ (the matrix with positive signature is
denoted as $\gamma_9$, ${\gamma_9}^2 ={\bf 1}$, while the eight
matrices with negative signatures are denoted as $\gamma_j$,
$j=1,2\ldots , 8$, with ${\gamma_j}^2 =-{\bf 1}$).  We are now in
the position to explicitly construct the whole series of primitive
maximal Clifford algebras $C(0,3+8n)$, $C(0,7+8n)$ through the
formulas
\begin{eqnarray}
C(0,3+8n)&\equiv& \begin{array}{lcr} {\overline\tau}_i\otimes
\gamma_9\otimes \ldots&\ldots&\ldots\otimes\gamma_9,\\ {\bf
1}_4\otimes\gamma_j\otimes{\bf 1}_{16}\otimes\ldots & \ldots &
\ldots\otimes{\bf 1}_{16},\\
 {\bf 1}_4\otimes\gamma_9\otimes\gamma_j\otimes {\bf
1}_{16}\otimes\ldots &\ldots&\ldots\otimes{\bf 1}_{16},  \\ {\bf
1}_4\otimes\gamma_9\otimes\gamma_9\otimes\gamma_j\otimes {\bf
1}_{16}\otimes \ldots&\ldots&\ldots\otimes{\bf 1}_{16},  \\ \ldots
&\ldots&\ldots, \\ {\bf
1}_4\otimes\gamma_9\otimes\ldots&\ldots&\otimes
\gamma_9\otimes\gamma_j,
\end{array}\label{quatern}
\end{eqnarray}
and similarly
\begin{eqnarray}
C(0,7+8n)&\equiv& \begin{array}{lcr} {\tilde\tau}_i\otimes
\gamma_9\otimes \ldots&\ldots&\ldots\otimes\gamma_9,\\ {\bf
1}_8\otimes\gamma_j\otimes{\bf 1}_{16}\otimes\ldots & \ldots &
\ldots\otimes{\bf 1}_{16},\\
 {\bf 1}_8\otimes\gamma_9\otimes\gamma_j\otimes {\bf
1}_{16}\otimes\ldots &\ldots&\ldots\otimes{\bf 1}_{16},  \\ {\bf
1}_8\otimes\gamma_9\otimes\gamma_9\otimes\gamma_j\otimes {\bf
1}_{16}\otimes \ldots&\ldots&\ldots\otimes{\bf 1}_{16},  \\ \ldots
&\ldots&\ldots, \\ {\bf
1}_8\otimes\gamma_9\otimes\ldots&\ldots&\otimes
\gamma_9\otimes\gamma_j,\label{octon}
\end{array}
\end{eqnarray}
Please notice that the tensor product of the $16$-dimensional
representation is taken $n$ times. The total size of the
(\ref{quatern}) matrix representations is then $4\times 16^n$,
while the total size of (\ref{octon}) is $8\times 16^n$.
\par
The formulas given above provide quite a practical and efficient
tool to operatively construct the irreducible Clifford algebras.
\par It should be noticed that all Clifford matrices are
even-dimensional (power of 2). An important subclass of Clifford
Gamma matrices is obtained by the matrices which are decomposable
in $2\times 2$ blocks and are non-vanishing only in the
anti-diagonal blocks. Such matrices can be named as (generalized)
Weyl-type matrices (they can also be regarded of ``supersymmetric
type" since they can be promoted to be fermionic matrices
associated with the representations of the extended
supersymmetries, see \cite{top}). An inspection of the previous
tables shows that the set of the (generalized) Weyl matrices is
found in special signature dimensions. All primitive Clifford
algebras are not of (generalized) Weyl type. However, all the
derived Clifford algebras, through the two lifting algorithms, are
of Weyl-type, once deleted the $ \left(
\begin{array}{cc}
   {\bf 1}_d &0 \\
  0 & -{\bf 1}_d
\end{array}\right)$ matrix to produce a non-maximal Clifford
algebra.\par To give a concrete example, the two-dimensional
Euclidean space $(2,0)$ is not of Weyl type, while the
two-dimensional Minkowski spacetime $(1,1)$ is of Weyl type.
Indeed, the first one is obtained from the $(2,1)$ Clifford
algebra by deleting a space-type Gamma matrix, while the second
one is obtained from the same $(2,1)$ Clifford algebra by deleting
one of the two temporal-type Gamma matrices. Without loss of
generality this Gamma matrix can always be chosen to be given by $
\left( \begin{array}{cc}
  {\bf 1}_d & 0 \\
  0 &-{\bf 1}_d
\end{array}\right)$. \par
The importance of the Weyl realization of Clifford algebras is of
course related with the possibility of introducing a Weyl
projection for Dirac spinors. The commutator between Gamma
matrices, $\Sigma_{\mu\nu} =[\Gamma_\mu,\Gamma_\nu]$, can be
regarded as the generator of the Lorentz group corresponding to
the given signature space-time. The product of two Gamma matrices,
both admitting non-vanishing blocks only in the antidiagonal,
correponds to a $2\times 2$ block matrix whose only non-vanishing
components are in the diagonal blocks. Since both the Gamma
matrices, as well as the Lorentz generators $\Sigma_{\mu\nu}$, act
on spinors, the fact that the Lorentz generators are
block-diagonals means that we can consistently set, under these
conditions, equal to $0$ half of the components of the column
vector spinors (either the upper half or the lower half), to
produce the so-called Weyl spinor, admitting half of the degrees
of freedom expected by the original Dirac spinor.  This reduction
of the components can be operated acting on a Dirac spinor with a
projector $P_\pm$ ($P_\pm P_\mp =0$ and ${P_\pm}^2= P_\pm$), given
by \begin{equation} P_\pm = \frac{1}{2}({\bf 1}\pm {\overline
\Gamma}) \label{projector}
\end{equation}
where ${\overline \Gamma} =
 \left( \begin{array}{cc}
  0 & {\bf 1}_d \\
  -{\bf 1}_d & 0
\end{array}\right)$.\par
In even-dimensional space-times the matrix ${\overline\Gamma}$ is
always given by the product of all the other $\Gamma$ matrices (it
corresponds to $\Gamma_5$ when we specialize to the standard
$4$-dimensional Minkowski space-time). \par In order to construct
lagrangian terms which are scalar under Lorentz transformations
and are given by bilinear products of spinors, we need to
introduce the notion of barred spinors ${\overline \Psi}$, given
by $\Psi^T\cdot A$, where $T$ denotes transposition (remember that
in our conventions spinors are without loss of generality assumed
to be real) and $A$ is a matrix, given by the product of all
temporal Gamma matrices, i.e. the generalization of the
Minkowskian $4$-dimensional $\Gamma_0$ matrix.

\section{Division algebras and Clifford algebras.}

So far we have shown how to construct the irreducible
representations of Clifford algebras, and not yet elucidated their
relations with division algebras. Such a relation can be expressed
as follows. The three matrices appearing in $C(0,3)$ can also be
expressed in terms of the imaginary quaternions $\tau_i$
satisfying $\tau_i\cdot\tau_j= -\delta_{ij}
+\epsilon_{ijk}\tau_k$. As a consequence, the whole set of maximal
primitive Clifford algebras $C(0, 3+8n)$, as well as their maximal
descendants, can be represented as quaternionic-valued matrices,
acting on spinors, which have to be interpreted now as
quaternionic-valued column vectors.\par Similarly, there exists an
alternative realization for the Clifford algebra $C(0,7)$,
obtained by identifying the seven generators with the seven
imaginary octonions satisfying the algebraic relation
\begin{eqnarray}
\tau_i\cdot \tau_j &=& -\delta_{ij} + C_{ijk} \tau_{k},
\label{octonrel}
\end{eqnarray}
for $i,j,k = 1,\cdots,7$ and $C_{ijk}$ the totally antisymmetric
octonionic structure constants given by
\begin{eqnarray}
&C_{123}=C_{147}=C_{165}=C_{246}=C_{257}=C_{354}=C_{367}=1&
\end{eqnarray}
and vanishing otherwise. This octonionic realization of the
seven-dimensional Euclidean Clifford algebra will be denoted as
$C_{\bf O}(0,7)$. Due to the non-associative character of the
(\ref{octonrel}) octonionic product (the weaker condition of
alternativity is satisfied, see \cite{gk}), the octonionic
realization cannot be represented as an ordinary matrix product
and is therefore a distinct and inequivalent realization of this
Euclidean Clifford algebra with respect to the one previously
considered (\ref{c07}). Please notice that, by iteratively
applying the two lifting algorithms to $C_{\bf O}(0,7)$ we obtain
matrix realizations (with octonionic-valued entries) for the
maximal Clifford algebras of the series $C(n, 7+n)$ and $C(8+n,
n-1)$, for positive integral values of $n$ ($n=1,2,\ldots$). The
dimensionality of the corresponding octonionic-valued matrices is
$2^n\times 2^n$. For completeness we should point out that the
construction (\ref{octon}) leading to the primitive maximal
Clifford algebras $C(0, 7+8n)$, can be carried on with the help of
an octonionic-valued realization of the $\gamma_9$ matrix. As a
consequence, realizations of $C(0,7+8n)$ and their descendants can
be produced acting on column spinors, whose entries are tensor
products of octonions. In any case, in the following, we will
focus on the single octonionic realizations $C_{\bf O} (n, 7+n)$
and $C_{\bf O}(9+n, n)$ (here $n=0,1,2,\ldots $) which are of
relevance in the context of the $M$-theory.
\par
One should be aware of the properties of the non-associative
realizations of Clifford algebras. In the octonionic case the
commutators $\Sigma_{\mu\nu} =\relax [\Gamma_\mu, \Gamma_\nu]$ are
no longer the generators of the Lorentz group. They correspond
instead to the generators of the coset $SO(p,q)/G_2$, being $G_2$
the $14$-dimensional exceptional Lie algebra of automorphisms of
the octonions. As an example, in the Euclidean $7$-dimensional
case, these commutators give rise to $7=21-14$ generators,
isomorphic to the imaginary octonions. Indeed
\begin{eqnarray}
\relax [\tau_i,\tau_j]& = &2C_{ijk}\tau_k .\label{octcomm}
\end{eqnarray}
The alternativity property satisfied by the octonions implies that
the seven-dimensional commutator algebra among imaginary octonions
is not a Lie algebra, the Jacobi identity being replaced by a
weaker condition that endorses (\ref{octcomm}) with the structure
of a Malcev algebra (see \cite{gk}).\par Such an algebra admits a
nice geometrical interpretation \cite{{hl},{cp}}. Indeed the
normed $1$ unitary octonions $X=x_0 + x_i\tau_i$ (with $x_0$ and
$x_i$, for $i=1,\ldots,7$, real and the summation over repeated
indices understood), i.e. restricted by the condition
\begin{eqnarray}
X^\dagger\cdot X&=&1,\label{uninorm}
\end{eqnarray}
describe the seven-sphere $S^7$. The latter is a parallelizable
manifold with a quasi (due to the lack of associativity) group
structure. Here $X^\dagger$ denotes the principal conjugation for
the octonions, namely $X^\dagger = x_0 -x_i\tau_i$.\par On the
seven sphere, infinitesimal homogeneous transformations which play
the role of the Lorentz algebra can be introduced through
\begin{eqnarray}
\delta X &=& a\cdot X,\end{eqnarray} with $a$ an infinitesimal
constant octonion. The requirement of preserving the unitary norm
(\ref{uninorm}) implies the vanishing of the $a_0$ component, so
that $a \equiv a_i\tau_i$. Therefore, the above commutator algebra
(\ref{octcomm}), generated by the seven $\tau_i$, can be
interpreted as the algebra of ``quasi" Lorentz transformations
acting on the seven sphere $S^7$. At least in this specific
example we discovered a nice geometrical setting underlining the
use of the octonionic realization of the $C_{\bf O}(0,7)$ Clifford
algebra. While the associative (\ref{c07}) representation of the
seven dimensional Clifford algebra is required for describing the
Euclidean $7$-dimensional flat space, the non-associative
realization describes the geometry of $S^7$.

\section{Division algebras and generalized supersymmetries.}

It is clear that extra-conditions on the generalized
supersymmetries such as (\ref{Malg}) can be imposed if we assume
the fundamental spinors being division-algebra valued (over ${\bf
C}$, ${\bf H}$ or ${\bf O}$) and not just real. A division algebra
analog of the supertranslation algebra can be introduced through
the position
\begin{eqnarray}
&\{Q_a, Q_b\} = \{{Q^\dagger}_a, {Q^\dagger}_b\} =0,&\nonumber\\
&\{Q_a, {Q^\dagger}_b\} = Z_{ ab}.&\label{divalgsusy}
\end{eqnarray}
where $\dagger$ denotes the principal conjugation in the given
division algebra and, as a result, the bosonic abelian algebra on
the r.h.s. is constrained to be hermitian
\begin{eqnarray}
Z_{ab}& =& {Z_{ba}}^\dagger.
\end{eqnarray}
Division-algebra structures for spinors can be consistently
imposed only in specific signature space-times. As already
recalled, in $D=11$, either a real or an octonionic structure can
be defined for the Minkowskian $C(10,1)$ case, while a
quaternionic structure can be imposed for the Euclidean $C(0,11)$
Clifford algebra (from formula (\ref{quatern}), constructed in
terms of the quaternions). The $32$ real components spinors can be
re-expressed in $(10,1)$ as $4$-component octonionic-valued
spinors and, for $(0,11)$, as $8$-component quaternionic-valued
spinors. In the Minkowskian $(10,1)$ case, the hermiticity
condition imposed on the $4\times 4$ octonionic-valued hermitian
matrix $Z_{ab}$ leaves it with $52$ independent components, while
in the Euclidean $(0,11)$ case the same hermiticity condition,
applied this time on the $8\times 8$ quaternionic-valued $Z_{ab}$
matrix, leaves only $120$ surviving bosonic components. Not
surprisingly, in both cases this number is less than the total
number of $528$ independent components obtained from the real
structure. This is already a first indication of the constraint
produced by the division algebra structures (especially the
octonionic one).
\par
It is worth concluding this section with a comment concerning the
reconstruction of the division algebra-valued matrix realizations
of algebraic structures in terms of their component fields. This
is better illustrated with a specific example. Let us discuss the
simplest case, the $1$-dimensional octonionic $N=8$ supersymmetry
(the considerations below trivially applies to the quaternionic
$N=4$ supersymmetry as well).\par Let us specialize
(\ref{divalgsusy}) to the two one-dimensional octonionic-valued
fermionic operators ${\cal{Q}}$, ${\overline{\cal Q}}$, satisfying
the $N=8$ superalgebra
\begin{eqnarray}
\{ {\cal Q}, {\cal Q}\} =\{{\overline{\cal Q}},{\overline{\cal
Q}}\} =0, &\quad& \{{\cal Q}, {\overline{\cal Q}}\} =
{\cal{H}}.\label{susyn8}
\end{eqnarray}
where ${\cal H}=H $ is real-valued and represents the
hamiltonian.\par ${\cal{Q}}$, ${\overline{\cal Q}}= {\cal
Q}^\dagger$ contain a total number of $8$ components and we should
expect they define an algebra with a total number of $8+
\frac{8\times7}{2}=36$ anticommutation relations. On the other
hand, the r.h.s. of (\ref{susyn8}) provides us at most of $3\times
8 = 24$ relations, so that it looks like something is missing.
Furthermore, when expanding in terms of components ($i=1,\ldots,
7$)
\begin{eqnarray}
{\cal Q} &=& Q_0 +\sum_i Q_i t_i,\nonumber\\ {\overline{\cal Q}}
&=& Q_0 -\sum_i Q_i t_i
\end{eqnarray}
and taking into account the (\ref{octonrel}) algebraic relations
for imaginary octonions, we end up with the following set of
relations for the component fields $Q_0$, $Q_i$ (the convention
over repeated indices is understood)
\begin{eqnarray}
\{Q_0,Q_0\} -\{Q_i,Q_i\} &=&0,\nonumber\\ \{Q_0,Q_i\}
&=&0,\nonumber\\ C_{ijk} [Q_j,Q_k] &=&0,\nonumber\\ \{Q_0,Q_0\}
+\{Q_i,Q_i\} &=&H, \label{susyodd}
\end{eqnarray}
These relations are odd since, in particular, the third one
involves a commutator, instead of an anticommutator as we should
expect. However, there is nothing wrong with (\ref{susyodd}) and
this algebra can be re-expressed in terms of its component fields
when correctly interpreted. The correct interpretation for
(\ref{susyodd}) consists in setting
\begin{eqnarray}
Q_0 &=& q,\nonumber\\ Q_i &=& \frac{i\lambda}{\sqrt{7}} \quad
for\quad any \quad i=1,\ldots,7.
\end{eqnarray}
In particular this implies that the ordinary component ${\bf Q_i}$
fields are not extracted from the $Q_i$ coefficients of $t_i$,
rather they have to be identified with the product $\lambda t_i$
\begin{eqnarray}
{\bf Q_i} &\equiv& {\lambda}t_i.\label{identif}
\end{eqnarray}
With the above position the set of ``odd" relations
(\ref{susyodd}) is now reduced to the set of ordinary relations
\begin{eqnarray}
\{ q, \lambda\} &=& 0,\nonumber\\ \{q,q\}=-\{\lambda,\lambda\} &=&
\frac{1}{2}H.
\end{eqnarray}
It should be noticed that the fermionic operator $\lambda$ is
antihermitian ($\lambda =-{\lambda}^\dagger$) in order to provide
the consistent hermiticity condition on ${\bf Q_i}$. It is worth
mentioning that all these relations are explicitly realized in the
octonionic matricial $N=8$ supersymmetry algebra which can be
recovered from the octonionic $2\times 2$ realization of $C_{\bf
O}(9,0)$. We recall that this octonionic $N=8$-supersymmetry
\cite{top} is constructed with the set of the hermitian $2\times
2$ octonionic-valued matrices of Weyl type non-vanishing only in
the antidiagonal (i.e. the additional $\Gamma_9$ matrix in $C_{\bf
O}(9,0)$ is discarded), given by
\begin{eqnarray}
\left(
\begin{array}{cc}
0& t_i \\ - t_i& 0
\end{array} \right) \equiv t_i\cdot
\left(
\begin{array}{cc}
0& 1\\ - 1& 0
\end{array} \right), &\quad &
\left(
\begin{array}{cc}
0& 1 \\ 1& 0
\end{array} \right).
\end{eqnarray}
We can identify
\begin{eqnarray}
q &\equiv & \left(
\begin{array}{cc}
0& 1 \\ 1 & 0
\end{array} \right),\nonumber \\
\lambda &\equiv & \left(
\begin{array}{cc}
0& 1 \\ - 1 & 0
\end{array} \right),
\end{eqnarray}
which satisfy the correct properties.\par The reconstruction of
the division-algebra structures in terms of its component fields
is more elaborated in the more complicated examples discussed in
the following. Nevertheless, even in these cases, it can be
performed following the procedure here outlined.

\section{The octonionic $M$-algebra.}

The octonionic $M$-algebra \cite{lt1} is defined by specializing
(\ref{divalgsusy}) to the $(10,1)$ case. The needed
octonionic-valued Clifford matrices are immediately obtained with
the help of the lifting algorithm introduced in section {\bf 2}
(e.g through the procedure $(0,7)\rightarrow (9,0) \rightarrow
(10,1)$, while the $C$ matrix introduced below coincides in this
case with the unique space-like Gamma matrix). It must be said
that two equivalent ways exist of introducing the $M$-algebra,
either in terms of the $4$-component $D=11$ spinors, or in terms
of the Weyl spinors in $(10,2)$ dimensions (the latter
construction leads to the $F$-algebra interpretation). Here we
just limit ourselves to consider the first case.\par The only
non-vanishing (anti)-commutator of the octonionic $M$-algebra is
given by
\begin{eqnarray}
\{ Q_a, {Q^\dagger}_b\} &=& Z_{ab},
\end{eqnarray}
where, in this case, the $52$ independent components of the
hermitian $Z_{ab}$ matrix can be represented either as the $11+41$
bosonic generators entering
\begin{equation}\label{eq1}
  {\cal{Z}}_{ab} = P^\mu (C\Gamma^{}_\mu)_{ab} +
   Z^{\mu\nu}_{\bf{O}} (C\Gamma^{}_{\mu\nu})_{ab}
   ,
\end{equation}
or as the $52$ bosonic generators entering
\begin{equation}\label{eq2}
  {\cal{Z}}_{ab} =
    Z^{[\mu_1\ldots \mu_5]}_{\bf{O}}
    (C\Gamma^{}_{\mu_1 \ldots
    \mu_5})_{ab}\, .
\end{equation}
The reason for that lies in the fact that, unlike the real case,
the sectors individuated by (\ref{eq1}) and (\ref{eq2}) are not
independent. This is a consequence of the non-associativity of the
octonions. Indeed, one has to point out that, when multiplying
antisymmetric products of $k$ octonionic-valued matrices, a
certain number of them are redundant. For $k=2$, due to the $G_2$
automorphisms, $14$ such products have to be erased. In the
general case \cite{crt} a table can be produced. We write it down
for $D$ odd-dimensional spacetime octonionic realizations of
Clifford algebras. The case suitable for $M$-theory is recovered
for $D=11$. The following table can be easily constructed from the
$D=7$ results (which are easily computed), by taking into account
that out of the $D$ Gamma matrices, $7$ of them are
octonionic-valued, while the remaining $D-7$ are purely real.
\par
The following table, up to $D=13$, is easily obtained:
{{\begin{eqnarray}&
\begin{tabular}{|c|c|c|c|c|c|c|c|c|c|c|c|c|c|c|}\hline
&$0$&$1$&$2$&$3$&$4$&$5$&$6$&$7$&$8$&$9$&$10$&$11$&$12$&$13$\\
\hline $D=7$&$1$&$7$&$7$&$1$&$1$&$7$&$7$&$1$&&&&&&\\ \hline

$D=9$&$1$&$9$&$22$&$22$&$10$&$10$&$22$&$22$&$9$&$1$&&&&\\ \hline

$D=11$&$1$&$11$&$41$&$75$&$76$&$52$&$52$&$76$&$75$&$41$&$11$&$1$&&\\
\hline

$D=13$&$1$&$13$&$64$&$168$&$267$&$279$&$232$&$232$&$279$&$267$&$168$&$64$&$13$&1\\
\hline

\end{tabular}&\nonumber\end{eqnarray}}}
\begin{eqnarray}
&&
\end{eqnarray}
where the columns are labelled by $k$, the number of
antisymmetrized Gamma matrices.
\par
To reproduce the above formulas one has to be careful in defining
the antisymmetric product for $k>2$ octonionic $\Gamma$-matrices.
Due to the non-associativity of the octonions the order of
multiplications must be correctly specified. The correct
prescription is the following one. The antisymmetrized product of
$k$ octonionic matrices $A_i$ ($i=1,2,\dots, k$) is given by
\begin{eqnarray}
\relax [A_{1}\cdot A_{2}\cdot \dots \cdot A_k] &\equiv&
\frac{1}{k!}\sum_{perm.} (-1)^{\epsilon_{i_1\dots i_k}}
(A_{i_1}\cdot A_{i_2}\dots \cdot A_{i_k}), \label{antisym}
\end{eqnarray}
where $(A_1\cdot A_2\dots \cdot A_k)$ denotes the symmetric
product
\begin{eqnarray}
(A_1\cdot A_2 \cdot\dots  \cdot A_n) &\equiv& \frac{1}{2}(. ((A_1
A_2)A_3\dots )A_k) +\frac{1}{2} (A_1(A_2(\dots A_k)).).
\end{eqnarray}
This prescription is consistent and produces the correct result.
As an example, in $D=11$ the three-fold antisymmetric product of
octonionic $\Gamma$-matrices $\relax C[\Gamma_i\cdot \Gamma_j\cdot
\Gamma_k]$ furnishes the $75$ antihermitian matrices, appearing in
the table above, describing together with $C$ an arbitrary
$4\times 4$ octonionic antihermitian matrix. The definition
(\ref{antisym}), applied to the five-fold products of the $D=11$
octonionic $\Gamma$-matrices, provides their octonionic
hermiticity. The explicit computation above shows that precisely
$52$ independent real tensorial charges describes the five-tensor
sector of the octonionic M-algebra, which means that it spans the
arbitrary $4\times 4$ octonionic-hermitian matrices. We thus see
that one can equivalently write the octonionic M-algebra  as
(\ref{eq1}) or as (\ref{eq2}). In the latter case, out of the
$462$ real antisymmetric $5$-tensorial charges of the standard
M-algebra, only $52$ are linearly independent, due to the
discovered relation
\begin{eqnarray}
\relax [\Gamma_{{\mu_1}\dots\mu_5}] &=&
{A_{[\mu_1\dots\mu_5]}}^\nu \Gamma_\nu +
{A_{[\mu_1\dots\mu_5]}}^{[\nu_1\nu_2]}\Gamma_{[\nu_1}\Gamma_{\nu_2]},
\label{M2M5}
\end{eqnarray}
with constant $c$-number coefficients
${A_{[\mu_1\dots\mu_5]}}^\nu$,
${A_{[\mu_1\dots\mu_5]}}^{[\nu_1\nu_2]}$.\par

The octonionic equivalence of different sectors (which, at least
for some spacetimes, can be interpreted as branes sectors) can be
simbolically expressed, in different odd space-time dimensions,
according to the table {{\begin{eqnarray}&
\begin{tabular}{|c|c|}\hline

$D=7$& $M0\equiv M3$\\ \hline

$D=9$&$M1+M2\equiv M4$\\ \hline

$D=11$&$M1+M2\equiv M5$\\ \hline

$D=13$&$M2+M3\equiv M6$\\ \hline

$D=15$&$M3+M4\equiv M0+M7$\\ \hline

\end{tabular}&\label{tablem}\end{eqnarray}}}

In $D=11$ dimensions the relation between $M1+M2$ and $M5$ can be
made explicit as follows. The $11$ vectorial indices $\mu$ are
splitted into the $4$ real indices, labelled by $a,b,c,\ldots$ and
the $7$ octonionic indices labelled by $i,j,k,\ldots$. We get, on
one side, {{\begin{eqnarray}&
\begin{tabular}{cc}

$4$& $M1_a$\\

$7$&$M1_i$\\

$6$&$M1_{[ab]}$\\

$4\times 7= 28$&$M2_{[ai]}$\\

$7$& $ M2_{[ij]}\equiv M2_{i}$

\end{tabular}&\nonumber\end{eqnarray}}}

while, on the other side, {{\begin{eqnarray}&
\begin{tabular}{cc}

$7$& $M5_{[abcdi]} \equiv M5_i$\\

$4\times 7=28$&$M5_{[abcij]}\equiv M5_{[ai]}$\\

$6$&$M5_{[abijk]}\equiv M5_{[ab]}$\\

$4$&$M5_{[aijkl]}\equiv {M5}_a$\\

$7$& $ M5_{[ijklm]}\equiv {\tilde M5}_{i}$

\end{tabular}&\nonumber\end{eqnarray}}}
which shows the equivalence of the two sectors, as far as the
tensorial properties are concerned. Please notice that the correct
total number of $52$ independent components is recovered
\begin{eqnarray}
52 &=& 2\times 7 +28+6+4.
\end{eqnarray}

\section{The octonionic superconformal $M$-algebra.}

In this section the superconformal octonionic M-algebra is
introduced following \cite{lt2}.\par The conformal algebra of the
octonionic M-theory can be introduced adapting to the eleven
dimensions the procedure discussed in \cite{cs} for the $10$
dimensional case. It requires the identification of the conformal
algebra of the octonionic $D=11$ $M$-algebra with the generalized
Lorentz algebra in the $(11,2)$-dimensional space-time. In such a
space-time the octonionic Clifford's Gamma-matrices are
$8$-dimensional. The basis of the hermitian generators is given by
the $64$ antisymmetric two-tensors
$
C\Gamma_{[\mu_1\mu_2]}{\cal Z}^{\mu_1\mu_2} $ and the $168$
antisymmetric three tensors $ C\Gamma_{[\mu_1\mu_2\mu_3]}{\cal
Z}^{\mu_1\mu_2\mu_3} $ (or, equivalently, by the $232$
antisymmetric six-tensors $ C\Gamma_{[\mu_1\ldots \mu_6]}{\cal
Z}^{\mu_1\ldots\mu_6} $). This is already an indication that the
total number of generators in the conformal algebra is $232$. We
will show that this is the case.

According to \cite{cs}, the conformal algebra can be introduced as
the algebra of transformations leaving invariant the inner product
of Dirac's spinors. In $(11,2)$ this is given by $\psi^\dagger C
\eta$, where the matrix $C$, the analogous of the $\Gamma^0$,
given by the product of the two space-like Clifford's Gamma
matrices, is real-valued and totally antisymmetric. Therefore,the
conformal transformations are realized by the octonionic-valued
$8$-dimensional matrices ${\cal M}$ leaving $C$ invariant, i.e.
satisfying
\begin{eqnarray}
{\cal M}^\dagger C + C {\cal M} &=& 0.
\end{eqnarray}
This allows identifying the (quasi)-group of conformal
transformations with the (quasi-)group of symplectic
transformations. Indeed, under a simple change of variables, $C$
can be recasted to be of the form
\begin{eqnarray}
\Omega &=&\left(
\begin{array}{cc}
0& {\bf 1}_4 \\ - {\bf 1}_4 & 0
\end{array} \right).
\end{eqnarray}
The most general octonionic-valued matrix leaving invariant
$\Omega$ can be expressed through
\begin{eqnarray}
{\bf M} &=&\left(
\begin{array}{cc}
D& B \\ C & -D^\dagger
\end{array} \right),\label{confM}
\end{eqnarray}
where the $4\times 4$ octonionic matrices $B$, $C$ are hermitian
\begin{eqnarray}
B=B^\dagger,\quad &&\quad C=C^\dagger .\label{BCcond}
\end{eqnarray}
It is easily seen that the total number of independent components
in (\ref{confM}) is precisely $232$, as we expected from the
previous considerations.

It is worth noticing that the set of matrices ${\bf M}$ of
(\ref{confM}) type forms a closed algebraic structure under the
usual matrix commutation. Indeed $\relax [ {\bf M}, {\bf M}]
\subset {\bf M}$, endowing the structure of $Sp(8|{\bf O})$ to
${\bf M}$. For what concerns the supersymmetric extension of the
superconformal algebra, we have to accommodate the $64$ real
components (or $8$ octonionic) spinors of $(11,2)$ into a
supermatrix enlarging $Sp(8|{\bf O})$. This can be achieved as
follows. The two $4$-column octonionic spinors $\alpha$ and
$\beta$ can be accommodated into a supermatrix of the form
\begin{eqnarray}&&
\left(\begin{array}{c|cc}
  0 & -\beta^\dagger & \alpha^\dagger\\ \hline
  \alpha & 0& 0 \\
  \beta & 0 & 0
\end{array}\right)\label{fermionic}.
\end{eqnarray}
Under anticommutation, the lower bosonic diagonal block reduces to
$Sp(8|{\bf O})$. On the other hand, extra seven generators,
associated to the $1$-dimensional antihermitian matrix $A$
\begin{eqnarray}
A^\dagger &=& - A, \label{Acond}
\end{eqnarray}
i.e. representing the seven imaginary octonions, are obtained in
the upper bosonic diagonal block. Therefore, the generic bosonic
element is of the form
\begin{eqnarray}&&
\left(\begin{array}{c|cc}
  A & 0 & 0\\ \hline
  0 & D& B \\
  0& C & -D^\dagger
\end{array}\right)\label{bosonic},
\end{eqnarray}
with $A$, $B$ and $C$ satisfying (\ref{Acond}) and
(\ref{BCcond}).\par The closed superalgebraic structure, with
(\ref{fermionic}) as generic fermionic element and (\ref{bosonic})
as generic bosonic element, will be denoted as $OSp(1,8|{\bf O})$.
It is the superconformal algebra of the $M$-theory and admits a
total number of $239$ bosonic generators.

\section{Conclusions.}

The octonions are at the very heart of many exceptional structures
in mathematics. It is very well known, e.g., that they can be held
responsible for the existence of the $5$ exceptional Lie algebras.
Indeed, $G_2$ is the automorphism group of the octonions, while
$F_4$ is the automorphism group of the $3\times 3$
octonionic-valued hermitian matrices realizing the exceptional
$J_3({\bf O})$ Jordan algebra. $F_4$ and the remaining exceptional
Lie algebras ($E_6$, $E_7$, $E_8$) are recovered from the
so-called ``magic square Tit's construction" which associates a
Lie algebra to any given pair of division algebras, if one of
these algebras coincide with the octonionic algebra \cite{bs}.\par
There is a line of thought \cite{ram} suggesting that Nature
prefers exceptional structures. Following this line of thought, in
\cite{smo}, the already recalled exceptional Jordan algebra
$J_3({\bf O})$ was used to define a unique Chern-Simons type of
theory in the loop quantum gravity approach. In a different line
of research, octonionic structures were investigated in different
works \cite{{fm},{cs}} in application to the superstring
theory.\par In this talk we summarized the results recently
obtained in a series of works, especially \cite{{lt1},{lt2}}
concerning the possibility of introducing an octonionic structure
for the $M$-theory algebra. After briefly recalling the
classification of spinors and Clifford algebras in terms of
division algebras (and more specifically their octonionic
construction) we were able to introduce at first the octonionic
$M$-superalgebra and, later, its superconformal extension
presented in formulas (\ref{fermionic}), (\ref{bosonic}).\par The
features of the octonionic $M$-superalgebra are puzzling. It is
not at all surprising that it contains fewer bosonic generators,
$52$, w.r.t. the $528$ of the standard $M$-algebra (this is
expected, after all the imposition of an extra structure, such as
the complex, quaternionic or octonionic structure, puts a
constraint on a theory). What is really unexpected is the fact
that new conditions, not present in the standard $M$-theory, are
now found. These conditions, which can be symbolically represented
in table (\ref{tablem}), imply that the different brane-sectors
are no longer independent. The octonionic $5$-brane contains the
same degrees of freedom and is equivalent to the $M1$ and the $M2$
sectors. We can write this equivalence, symbolically, as $M5\equiv
M1+M2$. This result is indeed very intriguing. It implies that
quite non-trivial structures are found when investigated the
octonionic construction of the $M$-theory. It also raises some
questions, because it is not yet clear how should we interpret it
and which is its proper meaning. At least two different viewpoints
can be advocated. On one hand, sticking with the original defined
octonionic algebra, one should try to investigate its possible
quantum-mechanical consistency, understanding whether and to which
extent it is possible to adapt the procedure of \cite{gpr} to the
present situation. On the other hand, another possibility can be
contemplated. We have discussed at the end of section {\bf 3} that
the octonionic realization of the $7$-dimensional Euclidean
Clifford algebra is related with the geometry of the seven sphere
$S^7$. There is a possibility, which deserves being investigated,
that the octonionic description of the $M$-theory would correspond
to a particular compactification of the $11$-dimensional
$M$-theory down to $AdS_4\times S^7$. This compactification
corresponds to a natural solutions for the $11$-dimensional
supergravity \cite{dp}. If this would be the case, the relations
of equivalence found in the octonionic construction should find a
counterpart in the $AdS_4\times S^7$ special compactification
geometry. Needless to say, this possibility is currently under
investigation.\par We conclude with a last remark that perhaps
deserves to be mentioned. We introduced both the conformal and the
superconformal extensions of the original $M$-algebra. They are
respectively given by $Sp(8|{\bf O})$ and $OSp(1,8|{\bf O})$, see
formulas (\ref{fermionic}), (\ref{bosonic}). $Sp(8|{\bf O})$ is
outside the scheme of conformal algebras of a given Jordan algebra
(such as $Sp(4|{\bf O})$, $Sp(6|{\bf O})$, the latter the
conformal algebra of $J_3({\bf O})$), usually investigated in the
mathematical literature, see \cite{{kan},{sud}}. The reason for
that is the fact that the bosonic sector of the $M$-algebra is
given by $4\times 4$ octonionic-valued hermitian matrices, and the
maximal Jordan algebra of octonionic-valued hermitian matrices is
given by $3$-dimensional matrices. The construction of the
conformal (and superconformal) algebra, however, as we have
proven, can be carried on in this case as well and it finally
produces the closed and consistent algebraic structures that we
mentioned before.


\end{document}